\documentstyle[11pt,newpasp,twoside,epsf]{article}
\def\edcomment#1{\iffalse\marginpar{\raggedright\sl#1\/}\else\relax\fi}
\reversemarginpar

\newcommand{\etal}{{\sl et~al.}}

\newcommand{\iu}{{\sl IUE}}
\newcommand{\fuse}{{\sl FUSE}}
\newcommand{\hs}{{\sl HST}}

\newcommand{\tlus}{\mbox{$\rm {\sc tlusty}\:$}}

\newcommand{\syn}{\mbox{$\rm {\sc synspec}\:$}}

\newcommand{\teff}{\mbox{$T_{\rm eff}\:$}}
\begin{document}

\title
{Heavy Elements in DA White Dwarfs}

\author{M.A. Barstow, M.R. Burleigh, N.P. Bannister}  
\affil{Department of Physics and Astronomy, University of Leicester,
University Road, Leicester LE1 7RH, UK}
\author{J.B. Holberg}
\affil{Lunar and Planetary Laboratory, University of Arizona, Tucson, 
AZ 85721, USA}
\author{I. Hubeny}
\affil{Laboratory for Astronomy and Solar Physics, NASA/GSFC, Greenbelt,
Maryland, MD 20711 USA}
\author{F.C. Bruhweiler}
\affil{Institute for Astrophysics \& \ Computational Sciences (IACS),
Dept. of Physics,
The Catholic Univ. of America, Washington DC 20064, USA}
\author{R. Napiwotzki}
\affil{Dr. Remeis-Strenwarte, Sternwartstr. 7, D-96049 Bamberg, Germany}

\begin{abstract}

We present a series of systematic abundance measurements for
a group of hot DA white dwarfs in the temperature range $\approx
20,000-75,000$K, based on far-UV spectroscopy with STIS on \hs , \iu
\ and \fuse . Using our latest heavy element blanketed non-LTE
stellar atmosphere calculations we have addressed the heavy element
abundance patterns for the hottest stars for the first time, showing that
they are similar to objects like G191$-$B2B. The abundances observed in the 
cooler ($<50,000$K) white dwarfs are something of a mystery. Some of the
patterns (e.g. REJ1032) can be explained by self-consistent
levitation-diffusion calculations but there is then a serious difficulty
in understanding the appearance of the apparently pure H atmospheres.
We also report the detection of photospheric HeII in the atmosphere of
WD2218+706.

\end{abstract}

\section{Introduction}

The existence of two distinct groups of hot white dwarfs, having either
hydrogen-rich or helium-rich photospheres is now, qualitatively at least, understood to
arise from the number of times the progenitor star ascends the red giant
branch and the amount of H (and He) lost through the successive phases of
mass-loss. Consequently, it seems clear that each group descends from
their respective proposed progenitors, the H-rich and He-rich central
stars of planetary nebulae (CSPN). Nevertheless, there remain several
features of the white dwarf cooling sequence that cannot be readily
explained. For example, while the hottest H-rich DA white dwarfs
outnumber the He-rich DOs by a factor 7 (Fleming \etal \ 1986), the
relative number of H- and He-rich CSPN is only about 3:1. In addition, 
there exists a gap in the He-rich track between $\approx 45000$K and 30000K
between the hot DO and cooler DB white dwarfs, confirmed by a detailed
spectroscopic analysis of all the then known hot He-rich objects by
Dreizler and Werner (1996).

To understand white dwarf evolution, we need to know accurately several
physical parameters for each star. For example, a measurement of
effective temperature ($T_{eff}$) establishes how far along its
evolutionary sequence the star has progressed. 
A key result has been the establishment of the temperature scale for DA
stars through the determination of \teff \ from the Balmer line profiles
in optical spectra. The reliability of this work depends on the assumption
that the Balmer line profile technique is a reliable estimator of \teff \
in all cases. However, Barstow Holberg \& \ Hubeny (1998) have shown that
the presence of substantial blanketing from photospheric heavy elements
does significantly alter the Balmer line profiles at a given effective
temperature. Hence, the temperature scale of the hottest, most metal-rich
DA white dwarfs, realised by studies using only pure H photospheric
models, cannot be viewed as reliable and must be revised taking into
account the photospheric composition of each star. 

Until recently very few DA white dwarfs were known to have effective
temperatures above $\approx 70000$K. Therefore, the proposed direct
evolutionary link between H-rich CSPN and white dwarfs has hardly been
explored. In particular, there have been no measurements of the
photospheric composition of what might be termed super-hot DAs, to
distinguish them from the cooler ranges studied in detail. Using a new
grid of non-LTE, heavy element-rich model atmospheres and a combination
of \hs \ STIS and \fuse \ spectra, we have made the first measurements of
photospheric abundance for a sample of very hot DA white dwarfs. For
comparison, we have also re-examined the archive data of the cooler DA
stars.

\section{Observations}

All the new far-UV spectroscopic observations were obtained as part of a
joint \hs \ STIS (cycle 8) and \fuse \ (cycle 1) programme. The \fuse \
observations are discussed elsewhere in these proceedings (Barstow \etal
\ 2000), with examples of typical spectra. Our STIS data were all
obtained with the E140M grating and cover the wavelength range from 1150
to 1700\AA . Figure~1 shows sample regions of the STIS spectrum of
REJ0558$-$371 (for which we also have \fuse \ data), illustrating the
detection of NV, OV, SiIV and FeV. The STIS spectra are typically more
sensitive to Fe and Ni than the \fuse \ observations and N is not
detected by \fuse . On the other hand, P and S which are detected in the
\fuse \ range cannot be seen by STIS. Hence, the two instruments acquire
complementary information.

\begin{figure}
\plottwo{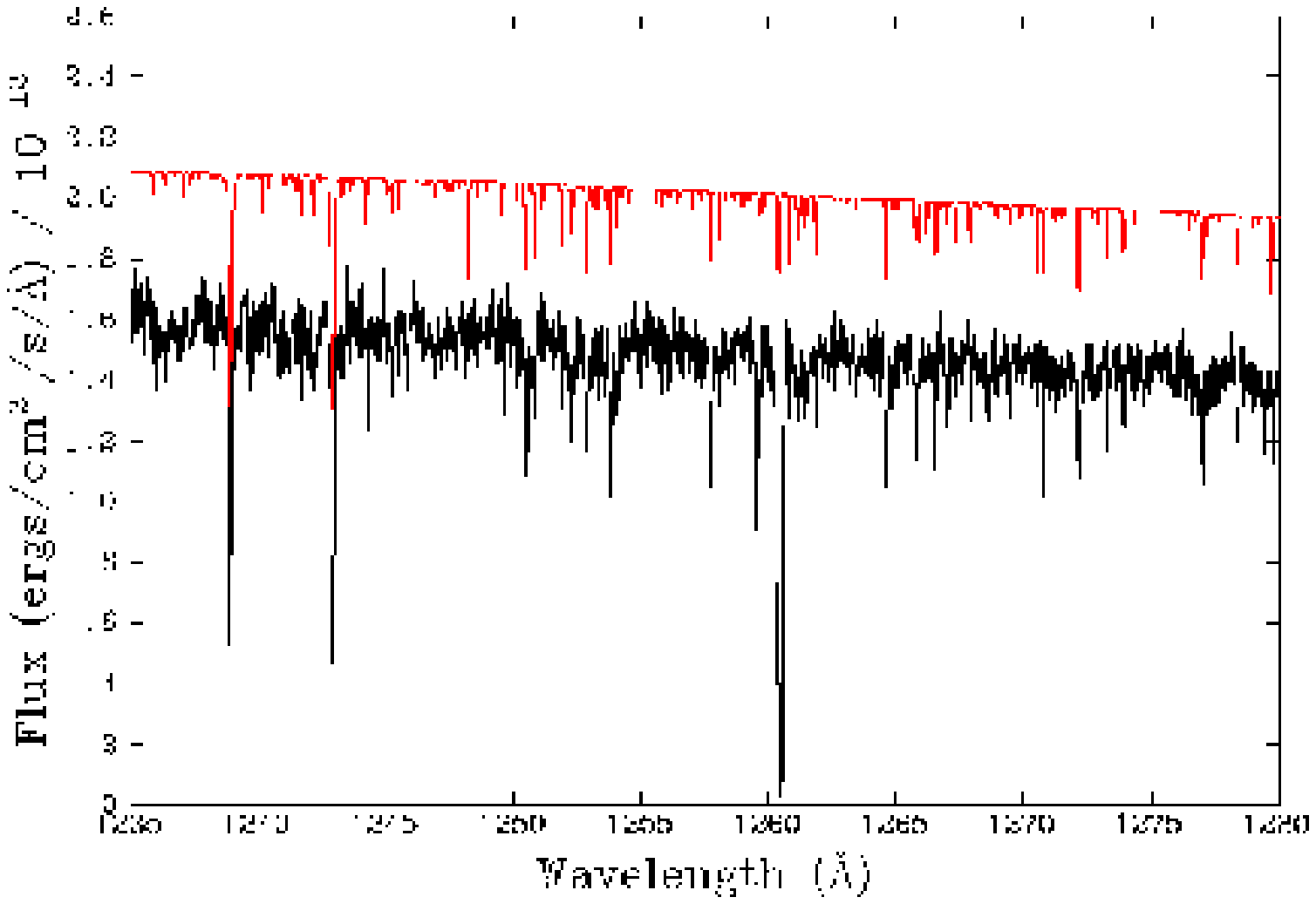}{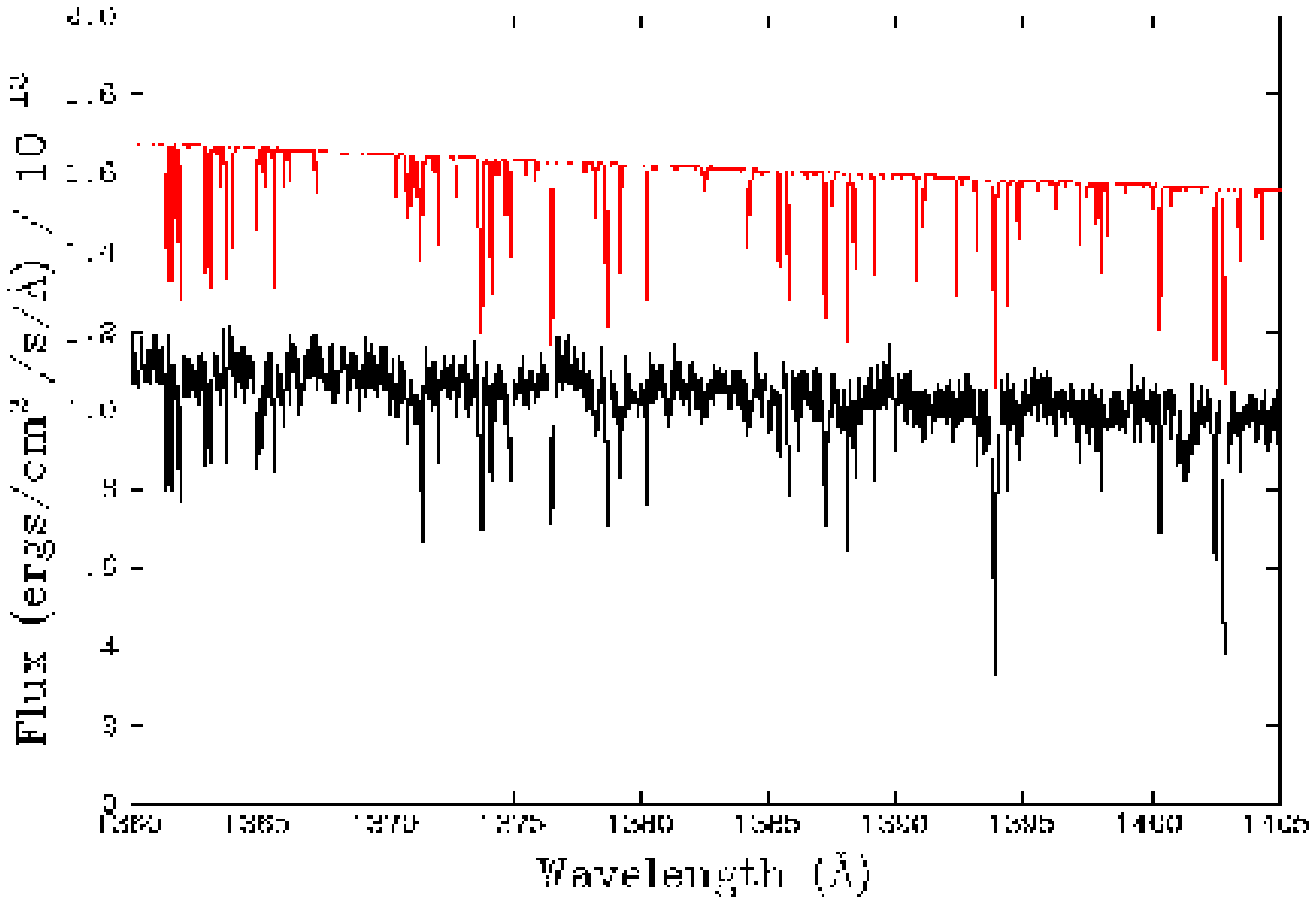}
\caption{Left: 1230\AA \ to 1280\AA \ region of the STIS spectrum of
REJ0558$-$371, showing photospheric absorption lines of NV
(1238.821/1242.804\AA ) and large
numbers of Ni lines. Right: 1360\AA \ to 1405\AA \ region of the
the STIS spectrum of REJ0558$-$371, showing photospheric lines of OV
(1371.296\AA ),
SiIV (1393.755/1402.770\AA ) and FeV. 
In both cases the best-fit synthetic spectrum is shown
offset for clarity.
}
\end{figure}

\section{Abundance measurements}

We have calculated a new grid of model stellar atmospheres using the
non-LTE code \tlus . These are based on work reported by Lanz \etal \
(1996) and Barstow, Hubeny \& \ Holberg (1998, 1999). In this case we
have extended the temperature range of the calculations up to 90,000K, to
deal with the hotter DA stars included in this analysis. To take account
of the higher element ionization stages that are likely to be encountered
in these objects, we have added new ions of OVI, FeVII/VIII and NiVII/VIII 
to the model atoms as well as extending the data for important ions such
as CIV to include more energy levels. As before, all the calculations
were performed in non-LTE with full line-blanketing. 
We initially fixed the abundances of the heavy elements at the values 
determined   from   our   earlier   homogeneous   analysis    of    G191$-$B2B 
(He/H=$1.0\times 10^{-5}$, C/H=$4.0\times 10^{-7}$,
N/H=$1.6\times 10^{-7}$, O/H=$9.6\times 10^{-7}$, Si/H=$3.0\times 10^{-7}$,
Fe/H=$1.0\times 10^{-5}$, Ni/H=$5.0\times 10^{-7}$). but taking into account that the CIV lines
near 1550\AA \ have subsquently been resolved into multiple component by
STIS. 

Abundances were estimated for each element in each star of the sample by
matching the data to a synthetic spectrum calculated from the \tlus \ non-LTE
model nearest in \teff \ and log g using \syn . Abundances were then varied
within a narrow range within \syn \ to obtain the formal values that give the
best representation of the data, summarised in Table~1. For the hot group
of stars (REJ0457 and above), gaps in the abundance data arise mostly from
the absence of data in the appropriate spectral range, rather than a true
absence of these elements. For GD246 and PG1123, which are of similar
temperature to REJ0457, the situation is not completely clear cut. STIS
observations are available for both these stars, but at higher dispersion
with the E140H grating and over a narrower wavelength range than the
E140M. Hence, the 1550\AA \ CIV resonance lines are not covered while NV,
SIV and OV are. For the cooler stars gaps in the table are indicative
that the element is not present at abundances detectable by STIS or \iu .

\begin{table}
\caption{Table 1. Photospheric heavy element abundances determined from
far-UV spectroscopy of the hot DA white dwarf sample.}
{\footnotesize 
\begin{tabular}{lllllllllll}
Star	&Teff&	log g&	C/H&	N/H&	O/H&	Si/H&	P/H&	S/H&	Fe/H&	Ni/H\\   
PG1342 & 72000 & 7.71 & 1.0E-5 &  & 2.0E-6 & 1.0E-6 & 1.0E-8 & 2.0E-6 & 1.0E-5 \\
REJ1738 & 71300 & 7.53 & 2.0E-8 & 3.0E-7 & 3.0E-7 & 1.0E-6 &  &  & 4.0E-6 & 5.0E-7 \\
REJ0558 & 63000 & 7.66 & 8.0E-7 & 3.0E-7 & 3.0E-6 & 2.0E-6 & 2.0E-8 & 2.5E-7 & 1.0E-5 & 1.5E-6 \\
REJ2214 & 62100 & 7.23 & 1.0E-6 & 7.5E-8 & 9.6E-7 & 7.5E-7 &  &  & 1.0E-5 & 1.0E-6 \\
REJ0623 & 59700 & 7.00 & 1.0E-6 & 1.6E-7 & 9.6E-7 & 3.0E-7 &  &   & 1.0E-5 & 1.0E-6 \\
WD2218 & 56900 & 7.00 & 4.0E-7 & 1.0E-6 & 1.0E-5 & 6.5E-7 &  &  & 2.0E-5 & 5.0E-7\\
Feige 24 & 56400 & 7.36 & 1.0E-7 & 3.0E-7 & 5.0E-7 & 3.0E-7 &  &  & 1.0E-5 & 2.0E-6 \\
REJ2334 & 54600 & 7.58 & 2.0E-8 & 5.0E-7 &   & 3.0E-7 &  &  & 1.0E-5 & 5.0E-7 \\
G191-B2B & 54000 & 7.39 & 4.0E-7 & 1.6E-7 & 9.6E-7 & 3.0E-7 & 2.5E-8 & 3.2E-7 & 1.0E-5 & 5.0E-7\\
GD246 & 53700 & 7.74 &   &   &   & 1.0E-7 &  &  &    &   \\
REJ0457 & 53600 & 7.80 & 4.0E-7 & 1.6E-7 & 9.6E-7 & 1.0E-7 & 2.5E-8 &  & 1.0E-5 & 5.0E-7 \\
PG1123 & 52700 & 7.52 &  &  &  &  &  &  &  &  \\
HZ43 & 49000 & 7.90 &   &   &   &   &   &   &   &  \\
REJ1032 & 46300 & 7.78 & 4.6E-7 & 5.0E-5 &   & 5.6E-8 &  &  &   &   \\
REJ2156 & 45900 & 7.74 &   &   &   &   &  &  &   &  \\
PG1057 & 39600 & 7.66 &   &   &   &   &   &  &   &   \\
REJ1614 & 38500 & 7.85 & 4.8E-7 & 2.5E-4 &   & 1.0E-8 &  &  &   &   \\
GD394 & 38400 & 7.84 &   &   &   & 8.0E-6 &  &  &  &  \\
GD153 & 37900 & 7.70 &   &   &   &   &  &  &   &   \\
GD659 & 35300 & 8.00 & 2.0E-7 & 6.3E-4 &   & 1.6E-8 &  &  &   &   \\
EG102 & 20200 & 7.90 &   &   &   & 1.0E-7 &  &  &   &   \\
Wolf1346 & 20000 & 7.90 &   &   &   & 3.2E-8 &  &  &   &  \\
\end{tabular}
}
\end{table}

\section{Discussion}

It is no surprise that significant quantities of heavy elements are
present in the majority of stars with \teff \ in excess of 50,000K.
What is particularly interesting for these hot objects is that the
abundances seen in the hottest object are quite similar to those seen in
the temperature regime near G191$-$B2B, the prototypical star of this
group. However, it is also clear that the abundances are not identical
from object to object. This variation may well be explained by small
differences in \teff \ and log g altering the precise balance of the
radiative levitation and diffusion processes. Schuh (2000)
has shown that including these effects self-consistently within the
model atmosphere calculations can give a good match to the observed
EUV and UV spectra of several of these stars. However, this agreement may
break down in the hotter objects if a wind is present and Schuh's analysis
should be extended to PG1342, REJ0558 and REJ1738.

The cooler white dwarfs (\teff $<50,000$K) can be divided into two broad
categories, those where heavy elements (typically C,N and Si, e.g.
REJ1032) are detected and those where the atmospheres seem to be devoid
of such material (e.g. REJ2156). Where heavy elements are detected in the
far UV, the EUV spectra appear to arise from a pure H envelope,
indicating that the atmospheres are highly stratified, with the heavy
elements residing in the outermost regions (see Holberg \etal \ 1995,
Holberg \etal \ 1999). In the case of one of these stars (REJ1032) Schuh
has demonstrated that the stratification (and resulting spectra) is a
natural consequence of the balance between radiative levitation and
downward gravitation. However, it is difficult to reconcile this picture
with the increasing abundance of N towards lower temperatures in REJ1614
and GD659. Furthermore, it is difficult to explain the dramatic
difference between objects at similar temperature and gravity, such as
REJ1032 and REJ2156 which has no heavy element content.

One or two objects appear to be anomalous in some way. While they fit
within the broad categories outlined above, the abundance of at least
one element is extreme compared to any other object. For example, the C
abundance measured from the \fuse \ spectrum of PG1342 is by far the
greatest value in the entire sample, by at least one order of magnitude.
Similarly GD394 contains a Si abundance well in excess of any other star.
In this case, Dupuis \etal \ (2000) have found the star to be peculiar in
other ways, exhibiting a 1.15d~ EUV photometric modulation. The data
suggest that GD394 is accreting material, but a possible source has not
yet been identified.

\section{Helium in WD2218+706}

All the above work deals with measured abundance of elements heavier than
H or He. However, the absence of He in the hot DAs remains a mystery.
CSPN typically have significant abundances of photospheric He but this
seems to disappear before the central star appears as a white dwarf.
One key aim of the STIS observations of the hottest white dwarfs was to
search for the presence of the HeII 1640\AA \ in the spectra. He was not
detected in either REJ1738 nor REJ0558, but is clearly present in
WD2218+706 (figure~2) and lies at the photospheric velocity. The implied
abundance of $3\times 10^{-5}$ is well below previous optical and UV
limits and less than predicted theoretically for \teff =56900K and log
g=6.9, appropriate for this star. It is tempting to mark this as the
first detection of photospheric He in an isolated hot DA (all other
detections appear to be associated with binary systems or possible merged
DAs). However, the gravity is rather low for an isolated star, suggesting
a binary origin, even though there is no companion known. In addition,
this star is associated with a known planetary nebula (DeHt5).

\begin{figure}[t!]
\plotone{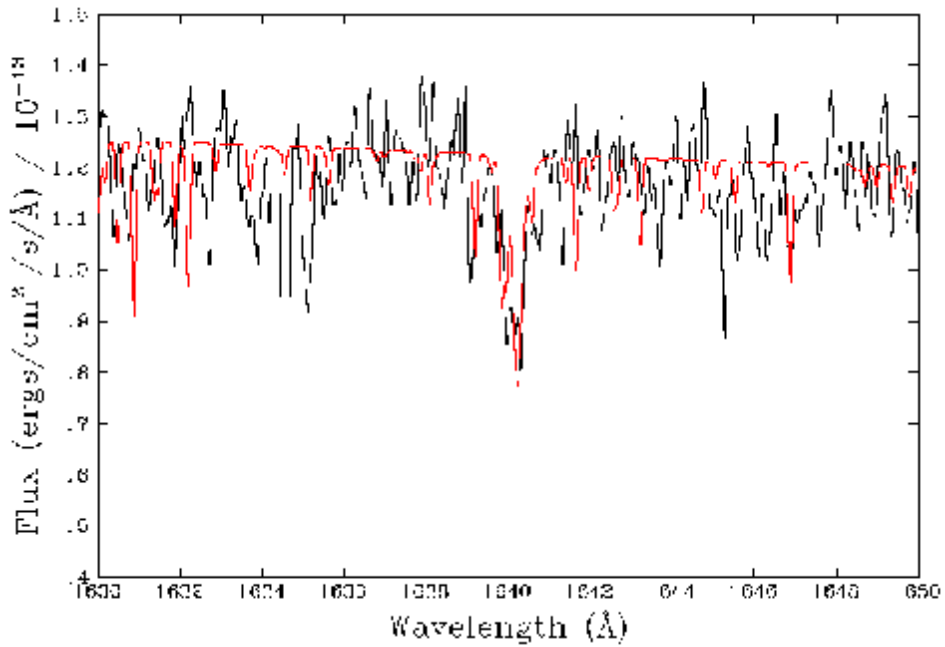}
\caption{STIS spectrum of WD2218+706 in the region of the HeII 1640\AA \
line, showing its clear detection. The smooth curve is a synthetic
spectrum calculated for an abundance (He/H) of $3\times 10^{-5}$.
}\vspace{-0.5cm}
\end{figure}

\section{Conclusion}

We have reported a series of systematic abundance measurements for
a group of hot DA white dwarfs in the temperature range $\approx
20,000-75,000$K, based on far-UV spectroscopy with STIS on \hs , \iu
\ and \fuse . Using our latest heavy element blanketed non-LTE
stellar atmosphere calculations we have addressed the heavy element
abundance patterns for the hottest stars for the first time, showing that
they are similar to objects like G191$-$B2B. The abundances observed in the 
cooler ($<50,000$K) white dwarfs are something of a mystery. Some of the
patterns (e.g. REJ1032) can be explained by self-consistent
levitation-diffusion calculations but there is then a serious difficulty
in understanding the appearance of the apparently pure H atmospheres.
New observations of other stars lying in this temperature range are need
to test the self-consistent model calculations.

\section*{Acknowledgements}

The work reported in the papers was based on observations
made with the \fuse \ and \hs \ observatories.
MAB, MRB and NPB were supported by PPARC, UK.

\end{document}